\documentclass[prl,aps,showpacs,twocolumn,floats]{revtex4} \input epsf
\usepackage{bm}
\begin{document}
\title{Effect of Intrinsic Curvature on Semiflexible Polymers}
\author{Surya K. Ghosh, Kulveer Singh and Anirban Sain}
\email{asain@phy.iitb.ac.in}
\affiliation{Physics Department,
Indian Institute of  Technology-Bombay, Powai, Mumbai, 400076, India. }

\begin{abstract}
Recently many important biopolymers have been found to possess intrinsic curvature.
Tubulin protofilaments in animal cells, FtsZ filaments in bacteria and double 
stranded DNA are examples. We examine how intrinsic curvature influence the 
conformational statistics of such polymers. We give exact results for the tangent-tangent 
spatial correlation function $C(r)=\langle\hat t(s).\hat t(s+r)\rangle$, both in two 
and three dimensions. Contrary to expectation, $C(r)$ does not show any oscillatory 
behaviour, rather decays exponentially and the effective persistence length has 
strong length dependence for short polymers. We also compute the distribution function
$P({\bf R})$ of the end to end distance ${\bf R}$ and show how curved chains can be 
distinguihed from WLC using loop formation probability. 
\end{abstract}

\maketitle

A cell hosts variety of polymeric filaments. Tubulin filaments in eukaryotic 
organisms and FtsZ filaments in prokaryotic organisms, are important members of 
the cellular cytoskeleton. They also play important role during cell division. 
Therefore in vitro properties of these filaments, both in their isolated filament 
form as well as in bundled form, have been of great interest \cite{molotsov,japEM,
ftsz-mica}. One of the important mechanical properties which tubulin and FtsZ 
filaments share is their intrinsic curvature. The mechanisms by which microtubules 
generate pulling force during chromosome segregation \cite{molotsov} and FtsZ-ring 
generates contractile radial force during cell constriction \cite{zring}, exploit 
their intrinsic curvature.

Interestingly biopolymers with intrinsic curvature and torsion have been known for 
long. Prominent examples are the alpha helices in proteins. They have also been 
extensively modeled \cite{yamakawa}. But polymers with only intrinsic curvature 
(but no intrinsic torsion), have not received much attention. This is perhaps 
because the helical structures are visually easy to detect compared to 
intrinsic curvature without torsion, except in two dimensions (2D). Intrinsic 
curvature is evident in electron microscope \cite{japEM} and and atomic force 
microscope (AFM)\cite{ftsz-mica} pictures of FtsZ and AFM pictures DNA 
\cite{curved-dna}. As our analysis will reveal, in three dimensions (3D), unless 
both $L/l_p$ and $l_p/R_0$ are of order one, where $R_0$ is the intrinsic radius 
of curvature of the polymer, $L$ its contour length and $l_p$ its persistence length, 
it is difficult to differentiate them, visually, from a WLC. In this letter we 
will focus on conformational properties of curved chains, that may distinguish it 
from a WLC. Such diagnostic markers are important in the context of ever increasing 
number of new biopolymers, with properties which cannot be explained by the standard 
models like Rouse, Zimm, freely jointed chain (FJC)\cite{Doi} or WLC, especially for short 
countour lengths. We ignore excluded volume effects and self crossings which are 
negligible if $L/R_0 \leq 1$; but important otherwise, as well as for non-dilute 
polymer solutions. Interplay of the additional length scale $R_0$ with $l_p$ and 
$L$ is of interest here.

Equilibrium, conformational statistics of such curved polymers have been studied using 
Monte-Carlo simulation \cite{macromol04} and auxiliary field theory \cite{MFT}. Their 
dynamics have been studied by numerical solution of their nonlinear equation of motion 
\cite{gold}. The field theoretic calculation \cite{MFT} assumed zero average torsion 
for the curved polymer and used a spatially varying auxiliary field. They predicted 
an oscillatory decay  $C(r)\sim e^{-r/\tilde l_p}\cos(r/R_0)$ for the correlation 
function in three dimensions. Here ${\tilde l_p}$ is the effective
persistence length. This oscillatory decay is geometrically suggestive, 
as the polymer may wind around itself (loop) due to its intrinsic curvature. 
This also implies that at large $l_p$, for $L>2\pi R_0$ the peak of the distribution 
function $P({\bf R})$ of the end to end distance $R$ will alternate between $R=0$ 
and $R=2R_0$, corresponding to complete loops and half loops, respectively,
as the length $L$ is increased.

We compute $C(r)$ exactly both in 2D and 3D and show that the decay is purely 
exponential, while oscillations can be recovered in 2D through extra constraint 
on the sign of the preferred inter-bond angle and through imposing torsion in 3D 
\cite{yamakawa}. The absence of oscillations can be understood as follows. 
Although intrinsic curvature constrains the magnitude of the local 
curvature $|\frac{d\hat t(s)}{ds}|\sim R_0^{-1}$, the direction of the curvature 
vector $\frac{d\hat t(s)}{ds}$ remain uncorrelated along the contour, leading
to exponential decorrelation. In other words the local plane of curvature keeps 
changing randomly which disfavors formation of planar or helical loops. Although 
the functional form of $C(r)$ cannot distinguish WLC from curved chains, the 
effective persistence length $\tilde l_p$ turns out to be $N$ dependent. We also 
compute the distribution function $P({\bf R})$ for curved chains which shows 
difference with WLC.

\begin{figure}
\centerline{\epsfxsize=11pc\epsfbox{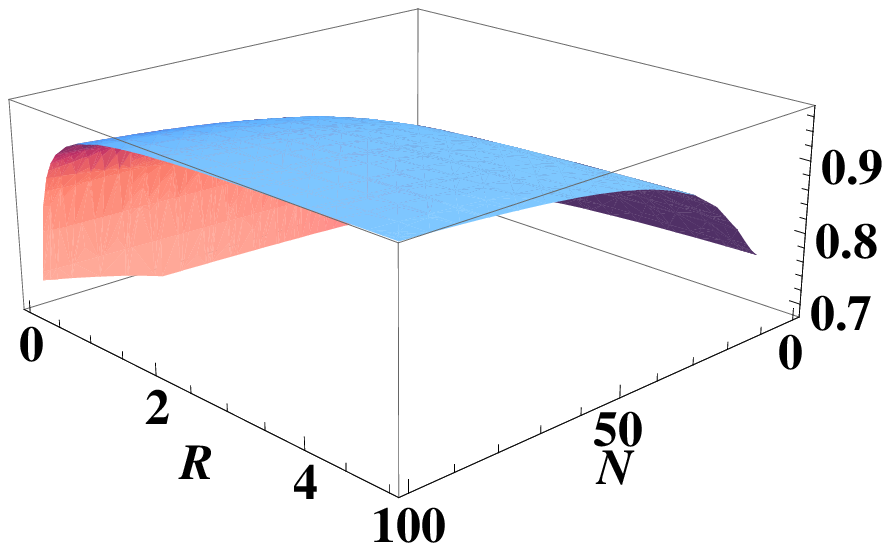} \hfill\hspace{-1pc} 
\epsfxsize=11pc\epsfbox{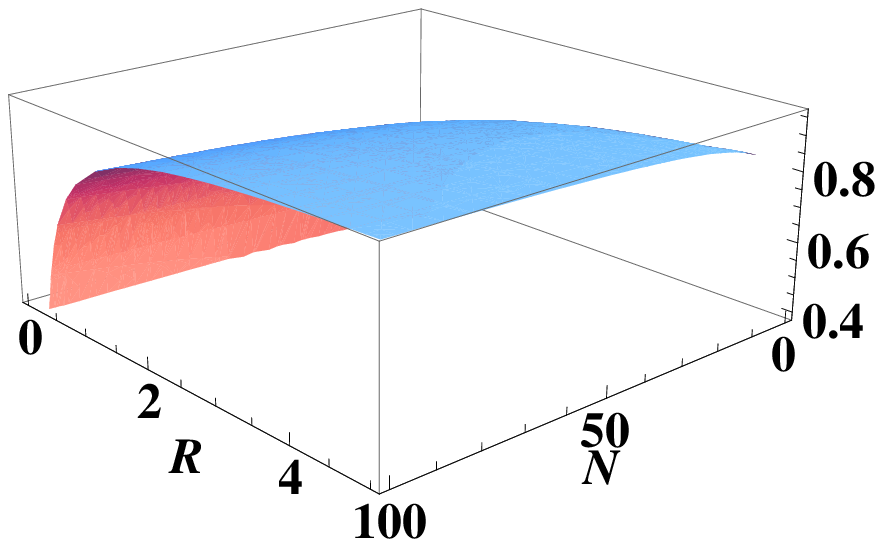} }
\caption{The effective persistence length $\tilde l_p/l_p$ with respect to $R_0$ and $N$,
at $l_p/L=0.2$ (left) and $l_p/L=10$ (right).} 
\label{fig.lpeff}
\end{figure}

A polymer of length $L$ is described by a space curve ${\bf R}  \left ( s \right ), 
s \in[0,L]$. The Hamiltonian of an intrinsically curved polymer is given 
by \cite {gold}, $\frac {H} {{K_B}T}= 
\frac {l_p} {2}  \int_{s=0}^{L} ~\left[\kappa(s) - R_0^{-1}\right] ^2 ds$. 
The curvature vector ${\vec\kappa}(s)=\frac{d\hat t(s)}{ds}$, where $\hat t(s)$
is the local tangent to the curve $(\hat t(s)=\;\frac{d{\bf R}(s)}{ds})$ and
$R_0$ is the intrinsic radius of curvature.  Changing variable $ x= \frac {s} {l_p} $, 
$\frac {H} {{K_B}T}= \frac {1} {2}  \int_{x=0}^{L/l_p}  ~
\left ( \left | \frac {d {\hat t}
\left (x\right)}{dx} \right | - \frac {l_p} {R_0} \right ) ^2 dx\;.
\label{eq.hamiltonian}$
This reveals that $l_p/L$ and $l_p/R_0$ are the two important 
dimensionless ratios of the problem. 
In the discrete limit the monomers are given by the position vectors ${\bf R}_i$ 
and the normalised bond vectors are 
$~\hat {t_i}=({\bf R}_{i+1} - {\bf R}_i)/b$, where $b$ is the 
bond length. The discretized Hamiltonian reads
\begin{equation}
\frac {H} {{k_B}T}= \displaystyle\sum_{i=1}^{N-1}h_i =\frac {1}{2} \displaystyle\sum_{i=1}^{N-1}
\left \{ \frac {{ \left | \hat {t}_{i+1} - \hat {t}_{i} \right | }}
{\sqrt \Delta}-
\frac {l_p \sqrt \Delta} {R_0}\right \}^2
\label{eq.Hi} 
\end{equation}
where $\Delta=\frac {L/l_p}{N}$.
Since the hamiltonian depends only on nearest neighbor bond angles, we can use a 
standard property \cite{LandauStat1}, $\langle\hat t_i {\bf.}\hat t_k\rangle=
\langle\hat t_i {\bf.}\hat t_{i+1}\rangle ^{|k-i|}$.
This yields $\langle\hat t(s) {\bf.}\hat t(s+r)\rangle=\exp(-r/\tilde l_p)$, where 
the spatial separation $r=b|k-i|$. Thus quite counterintutively, in 3D, $C(r)$ does not 
show any oscillation, but decays as $\exp(-r/\tilde l_p)$. We identify the effective 
persistence length as $\tilde l_p=-b/\ln\langle\hat t_i {\bf.}\hat t_{i+1}\rangle$ and 
$\langle\hat t_i{\bf.}\hat t_{i+1}\rangle=
\frac{\int d\Omega_i\cos\theta _i e^{-h_i(\theta _i)} }
{\int d\Omega_i e^{-h_i(\theta _i)} }$. Here $\theta_i$ is the angle between 
$\hat t_i$ and $\hat t_{i+1}$ and the integration is done by holding $\hat t_i$ 
as the polar axis and integrating $\hat t_{i+1}$ over the solid angle $\Omega_i$.
The thermal weight is $\exp(-h_i)$, where 
$h_i(\theta_i)=[\frac{\sqrt{1-\cos\theta_i}}{\sqrt \Delta}-\frac{l_p\sqrt \Delta}
{\sqrt 2R_0}]^2$ (see Eq.\ref{eq.Hi}). $\langle\hat t_i{\bf.}\hat t_{i+1}\rangle$ 
is given by  
\begin{widetext}
\begin{eqnarray}
e^{-y^2}\Big[2\sqrt N R_0 \Big(-L^2 l_p-2 L N R_0^2+2 l_p N^2 R_0^2+ 
e^{\frac{2l_p}{R_0}-y_1^2} [L^2 l_p+2l_p N^2 R_0^2+2 L N R_0 (l_p+R_0)]\Big)-
e^{y^2}\sqrt{2\pi Ll_p} (L^2 l_p+ 3LNR_0^2\nonumber\\   
-2l_pN^2R_0^2) \{erf[y]-erf[y-y_1]\} \Big]\Big /
\left[2l_p N^2 R_0^2 \big(2\sqrt NR_0\;e^{-y^2-y_1^2} \{e^{y_1^2}-e^{\frac{2l_p}{R_0}}\}+
\sqrt{2\pi L l_p}\{erf[y]-erf[y-y_1]\}\big)\right]
\label{eq.corel}
\end{eqnarray}
\end{widetext}
where $y=\sqrt{Ll_p/2N}/R_0,\;y_1=\sqrt{2Nl_p/L}$. 
Using this expression we plot $\tilde l_p /l_p$ in Fig.\ref{fig.lpeff} as a function of 
$R_0$ and $N$. At finite $R_0,L$, but large $N$, $\tilde l_p /l_p\rightarrow 1$.
But interestingly the discreteness of the chain has practical importance. For example, 
DNA has a finite bond length approximately equal to the size of a base
pair (bp) and experiments with short DNA strands ($15-90$bp) have revealed \cite{shortDNA}
an apparent persistence length which is $3-4$ times lesser than the standard value 
$\sim 50nm\;(150bp)$. To explain this Ref\cite{shortDNA} has ivoked new physics at the 
scale of base pairs. Models with long range correlation in the intrinsic curvature 
disorder in base pairs has also been proposed \cite{curved-dna}. But our calculation 
shows that a simpler factor, such as an uniform curvature can lead to substantial decrease
in $\tilde l_p$. Fig.\ref{fig.lpeff} shows that for low $R_0$ and high $l_p$ the 
reduction in $\tilde l_p$ could be two folds.

Since most curved polymers have been detected on 2D substrates, 2D requires special 
attention. We will now compute $C(r)$ in 2D. Here the angle between the bond vectors 
$\hat t_i$ and $\hat t_k$, $\theta_{i,k}$ can be written as a sum of the intermediate 
angles between the  successive bond vectors. Denoting $k=i+l,$ 
$\theta_{i,i+l}=\sum_{j=i}^{k-1}\theta_{j,j+1}$.
Note that the sign of the angles are important here and $\theta_{j,j+1}$ is
defined as the angle the $\hat t_{j+1}$ vector has to rotate with respect to
the $\hat t_j$ vector, and thus $C(r)$ is given by \\
\begin{eqnarray}
\langle\cos\theta_{i,i+l}\rangle=
{\it Re}\langle e^{i\sum_{j=i}^{k-1}\theta_{j,j+1}}\rangle
= \it {Re} 
\big[
\frac 
{\int d\theta e^{i\theta} 
e^{-h(\theta)} }
{\int d\theta 
e^{-h(\theta)} }
\big]^l \label{eq.2d}
\end{eqnarray}
Note that if the symmetry of $h(\theta)$  with respect to $\pm\theta$ is not broken then the
integrals $\int_{-\pi}^{\pi}$ are real and hence $\langle\cos\theta_{i,i+l}\rangle$
will decay exponentially, as in 3D. The integral can be estimated by noting that the 
numerator gets its maximum contribution from $\theta =\theta_0$ (and not $-\theta_0$, 
to break the symmetry) where the exponent $[\frac{\sqrt{1-\cos\theta_0}}
{\sqrt\Delta} -\frac{l_p\sqrt \Delta}{\sqrt 2 R_0}]$ is zero. In the continuum limit 
($b/R_o\ll 1$) the maximum occurs at a small value of $\theta_0\simeq b/R_0$ which allows 
the approximation $\cos\theta\simeq 1-\frac{\theta^2}{2}$. The resulting integral can 
be evaluated analytically (Eq.\ref{eq.Cs.2D}) and its agreement with the numerical
evaluation of the exact integral in Eq.\ref{eq.2d} is excellent (figure not shown).
\begin{widetext}
\begin{equation}
\it {Re}\{\big[
\frac
{\int_{-\pi}^{\pi} e^{i\theta} e^{-\frac{y_1^2}{2}(\theta-\theta_0)^2} d\theta }
{\int_{-\pi}^{\pi} e^{-\frac{y_1^2}{2}(\theta-\theta_0)^2} d\theta}
 \big]^l\}
=e^{-r/l_p} \cos(r/R_o)
\it {Re} \{\big[  
\frac
{ erf [ (i+y_1^2(\pi+\theta_0))/ \sqrt {2}y_1 ] - erf [ (i-y_1^2(\pi-\theta_0))/ \sqrt 2 y_1] }
{ erf[(\pi +\theta_0)\frac{y_1}{\sqrt 2}] +erf[(\pi -\theta_0) \frac{y_1}{\sqrt 2} ] }
\big]^l \}, 
\label{eq.Cs.2D}
\end{equation}
\end{widetext}
where $r=l(L/N)$. For finite $R_0,L$, but large $N$, the $N$-dependent factor
$\it {Re}[..]\rightarrow 1$ and we get $C(r)= e^{-r/l_p}\cos(r/R_o)$. 
Thus because of the preferred direction of 
the angle $\theta_0$ (as opposed to $\pm\theta_0$) the decay is oscillatory.
This is analogous to imposing an intrinsic torsion in 3D.

\begin{figure}[htbp]
\epsfxsize=8cm
\centerline{\epsfbox{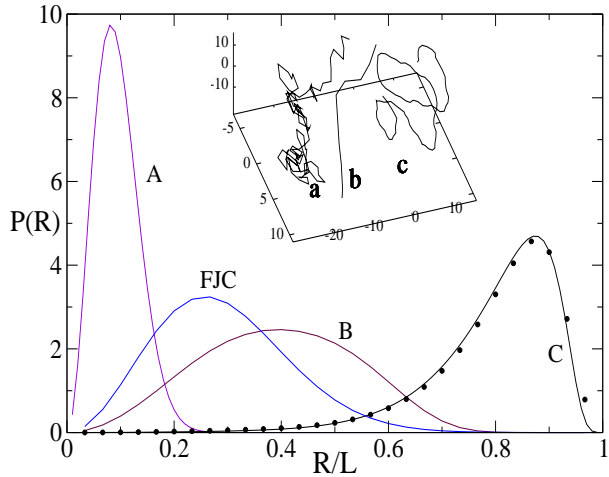}}
\vspace{0.2cm}
\caption{$P(R)$ vs $R/L$ for stiff chains ($L/l_p=1$) for different radii of curvatures, 
A) $R_0=0.1$, B) $R_0=0.3$ and C) $R_0=10$ (in units of $l_p$). 
At large $R_0/l_p$ the WLC limit (symbols) is 
reached while at small $R_0$ the peak of $P(R)$ can go past the flexible limit to smaller 
values of $R$, and thus can attain a globule size smaller than that of flexible Freely 
jointed chain (FJC). 
We used the same $N$ for all the chains.  Inset: Typical configurations of a) FJC, 
b) WLC and c) intrinsically curved chain. (b) and (c), both have $l_p/L=1$ and (c) 
in addition has $R_0=0.2$. 
These configurations were generated using MC simulation.}
\label{fig.big_Lp}
\end{figure}

\vspace{1cm}
\begin{figure}
\epsfxsize=7.5cm
\centerline{\epsfbox{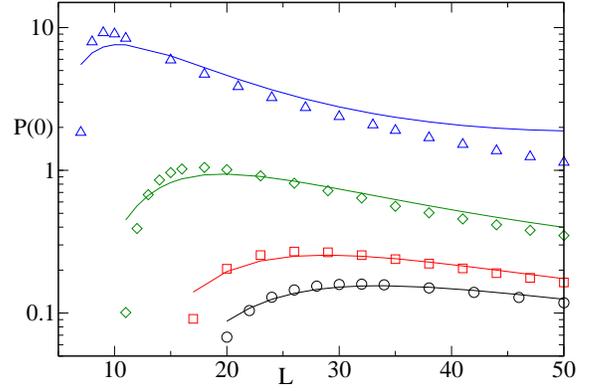}}
\vspace{.2cm}
\caption{
Probability of forming closed loops $P({\bf R}=0)$ versus contour length. From top to
bottom $R=1.5,3,7.5$ and $20$, at fixed $l_p=10$. At small $l_p/R_o$ (bottom-most) WLC 
is recovered. We fitted all the curves (solid lines) with the approximate formula 
derived by Shimada and Yamakawa \cite{yamakawa}: 
$P({\bf R}=0)=(896.32/l)^5\exp[-14.054/l+0.246l]$, for WLC in the range $l=L/l_p<10$. 
Clearly for large $l_p/R$ (the top two plots) the fit fails. That provides us a 
diagnostic tool for distinguishing a curved polymer from a WLC. }
\label{fig.4}
\end{figure}

We now compute the distribution function $P({\bf R})$, using transfer matrix methods. 
We also verified these results using a Monte-Carlo (MC) simulation which gave additional 
insights into the typical configurations of the curved chain (see inset of 
Fig.\ref{fig.big_Lp}). The distribution function of the end to end vector $~\bf R$ is 
 $ P \left ( \bf R \right ) = C \int d\hat {t}_1 \cdots \int d\hat {t}_N ~
 e^{-\beta H} ~ \delta \left ( \displaystyle\sum_{i=1}^{N} \hat{t}_i
 - \bf R \right )\; $, where C is the normalization constant for 
$\int d {\bf R} P \left ( {\bf R} \right ) =1 $. Following Ref.\cite{sam}, 
$P({\bf R})$ can be connected to the partition function 
$Z(f)=\int d\hat{t}_1 \cdots \int d\hat{t}_N ~exp\left
[-\beta H + f \int_{x=0}^{L/Lp} \hat{t}_i^z ~ dx \right ]$, 
where $f$ is an external field (analogous to external magnetic field in
Ising model).  This connection is made via the reduced probability 
distribution $ p \left ( z \right ) = \int d{\bf R}~ P \left ( R \right )  
~\delta \left ( R_3 -z \right ) $, which is the probability that 
one end of the polymer is fixed at $\bf R=0 $, and the other end 
lies in a given z plane. Using the definition of $P({\bf R})$ it turns out
that the Laplace transform of $p(z)$,  $\tilde p  \left ( f \right )
 =\int_{-L}^L ~ dz~\exp \left (fz/l_p \right ) ~p \left ( z \right ) =
{Z \left ( f \right ) }/ {Z \left ( f=0 \right ) }$. For computation 
purpose we converted this Laplace transform to a Fourier transform by
choosing $f$ purely imaginary. 
Substituting the Hamiltonian in $ Z \left ( f \right )$ we get
$Z \left ( f \right ) = \sum_{\hat t_1}..\sum_{\hat t_N}  
\exp\big\{\sum_{i=1}^{N-1}[h_i + \frac {f \Delta}{2} (\hat t_i + \hat t_{i+1} ).\hat z] 
+ \frac {f \Delta}{2}(\hat t_1+ \hat t_N ).\hat z\big\}$.
$Z \left ( f \right )$  can be computed using transfer matrices \cite{menon}.

\begin{eqnarray}
Z \left ( f \right ) =\displaystyle\sum_{\hat{t}_1}  \displaystyle\sum_{
\hat{t}_N} \langle \hat {t}_1 |V^{N-1}|\hat {t}_N \rangle ~ \exp
\left \{ \frac {f\Delta}{2} \left ( \hat {t}_1 + \hat {t}_N  \right ) \cdot \hat z
\right \}
\label{zfnumerical}
\end {eqnarray}
where the elements of the transfer matrix $V$ are $\langle \hat {t}_i |V|\hat {t}_{i+1} \rangle 
= \exp[-h_i+\frac {f \Delta}{2}( \hat {t}_i +\hat {t}_{i+1} )  \cdot \hat z]$.  After 
computing $Z(f)/Z(f=0)$ numerically, we inverse Fourier transform it to obtain $p(z)$ and using 
the relation $P(R)=-\frac{1}{2\pi z}\frac{dp(z)}{dz}|_{z=R}$ (which is obtained in
Ref.\cite{sam} using tomographic method and the isotropy of $P(R)$) we obtain $P(R)$.

For semiflexible polymers with intrinsic curvature the energy of a configuration 
is proportional to $l_p$, at fixed $L$. Therefore at large $l_p$, little deviation 
of local $R$ from $R_0$ makes the energy cost for such fluctuations large. Such rare 
but important small angle fluctuations, which requires large sampling in a Monte-carlo 
simulation, can be efficiently captured by transfer matrix method. In this stiff limit 
the peak of the end-to-end distribution function $P(R)$ is pushed away from the 
entropy dominated Gaussian peak (see Fig.\ref{fig.big_Lp}) and forces the polymer 
to make circular arcs of radius $R_0$. But at small $R_0$, very small globules, 
smaller than even that of an FJC (of same N), can form (Fig.\ref{fig.big_Lp}).
Whereas for small value of $l_p$ the behaviour will be dominated by the 
flexible limit and the peak of $P(R)$ moves towards the gaussian peak.

In Fig.\ref{fig.4} we show how loop formation probability $P({\bf R}=0)$ 
changes with contour length, in different regimes of $l_p/R$. Ref\cite{macromol04} 
had reported similar result using a Monte-carlo sampling. They found that 
as $R_0$ increases the peak of $P({\bf R}=0)$ unexpectedly shifts to $L$ values 
lesser than $2\pi R_0$. This was argued as thermal softening of the polymer. But 
it is unclear why thermal fluctuations should favour larger 
bending over smaller bending (relative to curvature $R_0^{-1}$). We argue that 
actually the shift occurs due to the competition between two different peaks in 
$P({\bf R}=0)$, and it also gives us a diagnostic tool to differentiate a curved 
chain from a WLC. At small $l_p/R_0$ i.e., in the WLC limit $P({\bf R}=0)$ has a 
peak at approximately $L/l_p\sim 3$ which emerge from a competition of bending 
energy and entropy \cite{yamakawa}. Whereas for large $l_p/R_0$ the curved chain 
has a peak at $L=2\pi R_0$ which is driven by energetics of intrinsic curvature. 
Notice that for the upper most plot (in Fig.\ref{fig.4}) the peak is located at 
$L/2\pi R_0=1$, while in the lower most plot it is located at $L/l_p\sim3$.
We note that Ref\cite{macromol04} had fitted their data for $P({\bf R}=0)$ 
with a formula in Ref\cite{yamakawa}, which however was derived for a curved 
polymer with nonzero torsion. 
 
Our transfer matrix calculation is different from the usual one in that we 
can not use periodic boundary condition which allows one to approximate 
the partition function as $\lambda_{max}^N$, where $\lambda_{max}$ is the
largest eigen value of the transfer matrix $V$. This is because we work 
in the regime $l_p\sim L$ where the effect of the boundary conditions 
can propagate deep inside the chain.  In other words, our system size is
not larger than the correlation length and hence strong finite size effects 
are expected. In Eq.\ref{zfnumerical}, $Z(f)$ is obtained as a weighted 
sum over all the matrix elements of $V^{N-1}$. 
In order to evaluate $V^{N-1}$ numerically, the angular space 
$\Omega\equiv(\theta,\phi)$ for a bond vector $\hat t_i$ is divided into 
$n=n_1\times n_2$ bins. Thus the matrix $V$ has dimension $n\times n$.
Note that while obtaining $V^{N-1}$ in Eq.\ref{zfnumerical} a
matrix multiplication $V^2=\int d\hat t V|\hat t\rangle \langle \hat t|V$ is 
numerically implemented through a discrete sum: 
$[V^2]_{ik}=\int d\Omega_j V(\Omega_i,\Omega_j)V(\Omega_j,\Omega_k)=
c\sum_{j=1}^n V_{ij}\sin \theta_j\delta_{jl}V_{lk}=c[VSV]_{ik}$. Here
the matrix $S_{jl}=\delta_{jl}\sin\theta_j$, with no summation over $j$ intended, 
\cite{mezard} arose from $d\Omega_j=\sin \theta_jd\theta_jd\phi_j$ and 
$c=d\theta d\phi=(\pi/n_1)\times(2\pi/n_2)$.

Finally, in order to get $V^{N-1}$ we needed a multiplication of the type 
$VSVSVS\dots V$. Since for high $l_p$ and $R_0$, the main contribution 
to $Z(f)$ comes from configurations where the inter-bond angles are small (and even 
smaller if $N$ is increased), the angular discretization has to be dense enough 
to pick up the contribution from the small interbond angles. This is computationally 
expensive. So we chose length discretizations as $N-2=2^m$ so that using $\Theta(\log_2 N)$ 
number of matrix multiplications we could obtain the matrix $(VS)^{N-2}$.

In summary, we showed by exact calculation that both in 2D and 3D, intrinsically 
curved polymers give exponentially decaying tangent correlation, $C(r)$, as 
semiflexible polymers. But the apparent persistence length $\tilde l_p$ is
substantially lesser than $l_p$ for short chains, when the ratio  $l_p/R_0$ is 
large. Contrary to physical expectation $C(r)$ does not have oscillatory decay 
with period $L=2\pi R_0$, unless we impose torsion in 3D or fix the sign of the 
preferred interbond bending angle in 2D. 
Curved biofilaments of FtsZ or tubulin can be distinguished from WLCs' through
loop formation probabilities, which has not been measured yet for curved polymers.  

We thank the Department of Science and Technology, India for financial support.\\

\end{document}